\documentclass[]{aastex62}

\received{July, 2019}
\submitjournal{ApJ}

\shorttitle{Identifying TeV Source Candidates}
\shortauthors{Chiaro et al.}
\begin{document}

\title{Identifying TeV Source Candidates among \emph{Fermi}-LAT Unclassified Blazars}

\correspondingauthor{Graziano Chiaro}
\email{graziano.chiaro@inaf.it}

\author{G. Chiaro}
\affil{Institute of Space Astrophysics \& Cosmic Physics, INAF \\ Via Bassini 15, I-20133 Milano Italy}

\author{M. Meyer}
\affiliation{Kavli Institute for Particle Astrophysics and Cosmology, Dpt. of Physics, SLAC, \\ Stanford University, Stanford, California 94305, USA}

\author{M. Di Mauro}
\affiliation{NASA  Goddard Space Flight Center \\ Greenbelt, MD 20771 USA}

\author{D. Salvetti}
\affiliation{Institute of Space Astrophysics \& Cosmic Physics, INAF \\ Via Bassini 15, I-20133 Milano Italy}

\author{G. La Mura}
\affiliation{Lab. de Instrumentacao e Fisica Experimental de Particulas. LIP \\ Av. Gama Pinto 2, Lisboa, Portugal}

\author{D. J. Thompson}
\affiliation{NASA  Goddard Space Flight Center \\ Greenbelt, MD 20771 USA}

%% Note that the \and command from previous versions of AASTeX is now
%% depreciated in this version as it is no longer necessary. AASTeX 
%% automatically takes care of all commas and "and"s between authors names.

%% AASTeX 6.2 has the new \collaboration and \nocollaboration commands to
%% provide the collaboration status of a group of authors. These commands 
%% can be used either before or after the list of corresponding authors. The
%% argument for \collaboration is the collaboration identifier. Authors are
%% encouraged to surround collaboration identifiers with ()s. The 
%% \nocollaboration command takes no argument and exists to indicate that
%% the nearby authors are not part of surrounding collaborations.

%% Mark off the abstract in the ``abstract'' environment. 
\begin{abstract}

Blazars and in particular the subclass of high synchrotron peaked Active Galactic Nuclei are among the main targets for the present generation of Imaging Atmospheric Cherenkov Telescopes (IACTs) and will remain of great importance for very high-energy $\gamma$-ray science in the era of the  Cherenkov Telescope Array (CTA).
Observations by IACTs, which have relatively small fields of view ($\sim$ few degrees), are limited by viewing conditions; 
therefore, it is important to select the most promising targets in order to increase the number of detections.
The aim of this paper is to search for unclassified blazars among known $\gamma$-ray sources from the {\it Fermi} Large Area Telescope (LAT) third source catalog that are likely detectable with IACTs or CTA. We use an artificial neural network algorithm and updated analysis of {\it Fermi}-LAT data.
We found 80 $\gamma$-ray source candidates, and
%for these sources we study their light curves and, 
for the highest-confidence candidates, we calculate their potential detectability 
with IACTs and CTA based on an extrapolation of their energy spectra. Follow-up observations  of our source candidates could significantly increase the current TeV source population sample and could ultimately confirm the efficiency of our algorithm to select TeV sources. 
%It could also lead to a revision of the predicted number of sources that will be detected in the CTA extragalactic survey.

\end{abstract}

%% Keywords should appear after the \end{abstract} command. 
%% See the online documentation for the full list of available subject
%% keywords and the rules for their use.
\keywords{gamma rays --- 
blazars --- catalogs --- surveys}

%% From the front matter, we move on to the body of the paper.
%% Sections are demarcated by \section and \subsection, respectively.
%% Observe the use of the LaTeX \label
%% command after the \subsection to give a symbolic KEY to the
%% subsection for cross-referencing in a \ref command.
%% You can use LaTeX's \ref and \label commands to keep track of
%% cross-references to sections, equations, tables, and figures.
%% That way, if you change the order of any elements, LaTeX will
%% automatically renumber them.
%%
%% We recommend that authors also use the natbib \citep
%% and \citet commands to identify citations.  The citations are
%% tied to the reference list via symbolic KEYs. The KEY corresponds
%% to the KEY in the \bibitem in the reference list below. 

\section{Introduction} \label{sec:intro}

Blazars,  some of the most powerful Active Galactic Nuclei (AGNs), have a relativistic jet pointing toward the observer \citep[e.g.,][]{abdo01,fra} and show rapid variability and high optical and radio polarization.  Such objects are the most numerous class of extragalactic sources detected by TeV telescopes, the most sensitive of which are the Imaging Atmospheric Cherenkov Telescopes (IACTs) such as  MAGIC\footnote{\url{https://magic.mpp.mpg.de/}}, H.E.S.S.\footnote{\url{https://www.mpi-hd.mpg.de/hfm/HESS/}},  VERITAS\footnote{\url{https://veritas.sao.arizona.edu/}},  and the upcoming Cherenkov Telescope Array (CTA)\footnote{\url{https://www.cta-observatory.org/}}. Despite their high sensitivity, however, observations by current IACTs are limited by their fairly small fields of view ($\sim$ few degrees), weather conditions, the need for relatively dark night skies, and by a high background that requires fairly long observations. A source with a flux of $\sim$1$\%$
of the Crab Nebula flux  requires around 50 hours of observation time for a detection at 5$\sigma$.  IACTs typically take data for only about 1200 hours per year \citep{Naurois}.  
Those constraints provide a strong incentive to identify likely targets for IACT observations. 

Blazar Spectral Energy Distributions (SEDs) show two 
broad peaks in a $\nu f{_\nu}$ representation. The low-energy peak is attributed to synchrotron radiation, and the high-energy one is usually thought to be produced by inverse Compton radiation (IC) \citep[e.g.][]{Sikora}.  Based on the position of the  synchrotron peak ($\nu^{S}_{peak}$) in the SED, blazars are divided into three subclasses:   low-synchrotron-peaked (LSP,  with $\nu^{S}_{peak} \leq 10^{14}\,$Hz), intermediate-synchrotron-peaked (ISP,  with $10^{14}\, {\rm Hz} < \nu^{S}_{peak} \leq 10^{15}\,$Hz) and high-synchrotron-peaked (HSP, with $\nu^{S}_{peak} > 10^{15}\,$Hz) \citep{abdo01}. 
%In this study we refer to those three subclasses  as listed in the Third Catalogue of Active Galactic Nuclei detected by the {\it Fermi} LAT \citep[3LAC:][]{3lac}. Synchrotron-peak data  come from the visual inspection of each LAT blazar SED by a team of {\emph{SEDders}} \footnote{{\url{https://www.asdc.asi.it/ }}} who fitted the SEDs in order to obtain the peak frequency and the classification of the source.
HSPs, primarily BL Lac objects, represent the most numerous class of extragalactic TeV-energy sources. The TeVCat\footnote{\url{http://tevcat.uchicago.edu/}} is an online, interactive catalog for very-high-energy %(VHE energies, $E > 50$ GeV) 
(VHE energies, $E > 100$ GeV) $\gamma$-ray astronomy \citep{horan}.
This catalog reports 223 TeV sources as of this writing.  Among the 61 objects associated with blazars,  51 of them are HSPs and only 10 are LSP/ISP flat-spectrum radio quasars (FSRQs)\footnote{The rest of the sources  in TeVCat are Galactic sources or of unidentified nature.}.
 %The 2WHSP catalog resulted from a cross-match of a number of multi-wavelength surveys (in the radio, infrared, and X-ray bands) and applied selection criteria based on the radio to IR and IR to X-ray spectral slopes.

All-sky observations with the Large Area Telescope (LAT) on board the Fermi Gamma-ray Space Telescope (Fermi) \citep{FLAT} at GeV  energies offer opportunities to find such targets.  An example is the Third Catalog of Hard {\it Fermi}-LAT Sources \citep[3FHL:][]{3fhl}, which reports the locations and spectra of sources significantly detected in the 10 GeV - 2 TeV energy range during the first 7 years of the {\it Fermi} mission.
From the 3FHL  %TeV-telescope 
it is possible to select  TeV candidates by $\gamma$-ray flux and photon index.

An alternative approach to searching for TeV candidates is to find objects belonging to a class of sources likely to be seen at TeV energies. In the case of blazars, this can be done by identifying those objects whose synchrotron emission peaks at high frequencies. An example of this approach is the second {\it WISE} High Synchrotron Peak Catalog (2WHSP) \citep{2wHSP}, which is a list of HSP candidates based on  multi-frequency analysis of $\gamma$-ray source candidates with $|b| > 10^\circ$.%For extragalactic sources, this means identifying AGN that have a peak energy output at high optical frequencies.

%The present search for TeV  HSP source candidates is a third approach, using a two-step method: 
Here we present a third approach to search for TeV HSP source candidates. This method includes two steps: (1) we use $\gamma$-ray variability information to search out potential HSPs among the unclassified {\it Fermi}-LAT sources; and (2) we analyze $\gamma$-ray spectra of these sources using more {\it Fermi}-LAT data than used in published catalogs.
%, in order to determine which of these HSP candidates are most likely to be detectable at TeV energies.  
The starting point is the third {\it Fermi}-LAT all-sky catalog  of sources detected at energies between 100 MeV and 300 GeV \citep[3FGL:][]{3FGL}.
The 3FGL catalog lists 3033 $\gamma$-ray sources, of which 1745 are AGNs, mostly BL Lacs and FSRQs, and includes $\gamma$-ray source locations, energy spectra, variability information on monthly time scales, and likely associations with objects seen at other wavelengths. In this catalog 573 sources are listed as  blazars of uncertain type (BCUs) and 1010 objects lack  a plausible counterpart at other wavelengths (Unassociated Catalog Sources, UCSs)\footnote{A preliminary version of the 4FGL catalog is available at \url{https://fermi.gsfc.nasa.gov/ssc/data/access/lat/8yr_catalog/}, but it does not include the variability information needed for this analysis.}.

Although BCUs and UCSs often lack optical spectra and sufficient information for a rigorous classification,  statistical methods such as the Artificial Neural Network (ANN) algorithm can potentially provide  classifications of these sources \citep[e.g.,][]{bflap,pablo, zoo, leu}. In particular,  \citet{pablo} found 559 of the UCS sources have characteristics similar to those of AGN.  These UCS$_{agns}$ are combined with the original 573 BCUs  from the 3FGL catalog to provide the targets for our search  %{\bf\textcolor{blue}{The Artificial Neural Network (ANN) algorithm is a common method for statistical studies of this type.}} %We focus on the uncertain types of blazars as ones less likely to have been studied by other methods.  %The Artificial Neural Network (ANN) algorithm is a common method for statistical studies of this type.  
%{\bf\textcolor{blue}{and we apply an  %{\textcolor{blue}{an optimized version of}} 
%the {\bf\textcolor{blue}{original}} 
%ANN algorithm 
%used by \citet{bflap} 
%to search for 
for HSP/TeV candidates.
%in order to select new extragalactic TeV-energy targets. \\

The paper is organized as follows: in Sect.~\ref{<2>} we present the machine learning method used in this study; in Sect.~\ref{<3>} we describe the selection of HSP candidates among the uncertain 3FGL objects; and in Sect.~\ref{<4>}  we discuss the results of a dedicated {\it{Fermi}}-LAT analysis of the sources found analyzing 104 months of data. In Sect.~\ref{<5>} we examine the detectability of the targets by the present generation of IACTs and CTA. In Sect.~\ref{<6>} we summarize the conclusions of this study.

\section{The ANN  search method}
\label{<2>}
The starting point for selecting HSP candidates is the ANN method previously applied to {\it Fermi}-LAT sources to distinguish FSRQ-like sources from those with BL Lac characteristics \citep{bflap, zoo}.  The key idea is that the $\gamma$-ray flares of BL Lacs tend to be smaller and less frequent than those of FSRQs. The input is the empirical cumulative distribution function (ECDF) of monthly $\gamma$-ray flux values for each source, taken from the 3FGL. We included in the algorithm values corresponding to increments of 10\% from the 10th to the 100th percentile of the ECDF. The  algorithm computes a likelihood value arranged to have two possibilities: class $A$ or class $B$, with a likelihood ({\it{L}}) assigned to each analyzed source so that the likelihoods to belong to one or the other of the two classes are related by $L_A = 1-L_B$. In this way, the greater the value of {\it{L}$_{A}$}, the greater the likelihood that the source is a class $A$ candidate. In this case, the likelihood applies to the source having HSP characteristics, $L_{HSP}$. This approach uses the two-layer-perceptron ANN technique \citep{gis90,bio}, which is probably the most widely used architecture for practical applications of neural networks.  Data enter the neural network through  nodes in the input layer. The
information travels across the links and is processed in the nodes through an activation function. Each node in
the output layer returns the likelihood of a source to be a specific
class.
%Fig.~\ref{per} shows a schematic view of a Two Layer Perceptron (2LP) ANN architecture we used for our analysis.
%\begin{figure}
%\includegraphics[width=0.5 \textwidth]{2LP.png}
%\caption{\label{per}{Schematic view of a Two Layer Perceptron (2LP), the
%Artificial Neural Network architecture we used for our analysis.
%Data enter the 2LP through the nodes in the input layer. The
%information travels from left to right across the links and is processed in the nodes through an activation function. Each node in
%the output layer returns the likelihood of a source to be a specific
%class}}
%\end{figure}
%Because of the relatively low complexity of our data, we use %{\textcolor{blue}{an optimized version of an}} 
%{\bf{the }} ANN algorithm described by \citet{bflap}, %{\textcolor{blue}{with}} 
%{\bf{that works by}} a 
%the simple structure known as Feed Forward multi-layer perceptron and in particular two-layer perceptron.
%In an ANN analysis, 
%Following the procedures of the original method \citep{bflap}, 
We applied the algorithm to the three  synchrotron peak subclasses as classified in the Third Catalog of Active Galactic Nuclei detected by the {\it Fermi} LAT \citep[3LAC:][]{3lac} in order to train it to distinguish each source class.

Repeating the analysis of \citet{bflap} as a cross-check on that analysis, we considered 289 HSPs and 824 non-HSP objects classified in the third {\it{Fermi}}-LAT AGN catalog \citep[3LAC;][]{3lac}. Maintaining the same ratio as in the catalog, that is, one third  HSPs and two thirds non-HSP sources,   we randomly mixed the sample and divided it  into 3 subsamples:  training,  validation, and testing. The training sample is used to optimize the network. The validation sample is used to avoid over-fitting. The testing sample is independent both of the training and validation ones and was used to monitor
the accuracy of the network.
Although the random sampling resulted in a different training set from the previous analysis, the results were the same:  for  $L_{HSP}$ $>$ 0.8, 75\% of the sources have characteristics of HSPs, while for $L_{HSP}$ $>$ 0.89, we expect 90\% of the sources to be HSP-like.
%during the training one third for training the algorithm, one third for optimization and remaining for testing respectively.
%Applying our algorithm to the %training 
%testing sample that corresponds to 15$\%$ of the full sample, the distribution reveals  %produces a likelihood distribution with 
%a concentration for non-HSP sources at low $L_{\emph{HSP}}$ and a relatively flat distribution for HSPs. Fig.~\ref{like}  shows that the likelihood distribution in the testing sample that we used in our study to reject non-HSP-type blazars. 
%Assessing the completeness of the sample and the fraction of spurious sources labelled as HSP-like or non HSP-like candidates we define two classification thresholds. The former is based on the optimization of the positive association rate (precision). It is defined as the fraction of true positives with respect to the objects classified as positive, of $\sim{90}\%$. The latter based on  the sensitivity, defined as the fraction of objects of a specific class correctly classified as such, referenced to the defined threshold, $L_{\emph{HSP}}$ $>$ 0.8.

%\begin{figure}
%\includegraphics[width=0.5 \textwidth]{fig/figure_6_sample3_HSP.png}
%\caption{\label{like}
%Distribution of the ANN  likelihood of  HSP ( blue) and non-HSP( red)  of 3 LAC  sources in the testing sample. } 
%\end{figure}

\section{Identifying HSP candidates} 
\label{<3>}

\begin{figure*}[!]
\includegraphics[width=0.5 \textwidth]{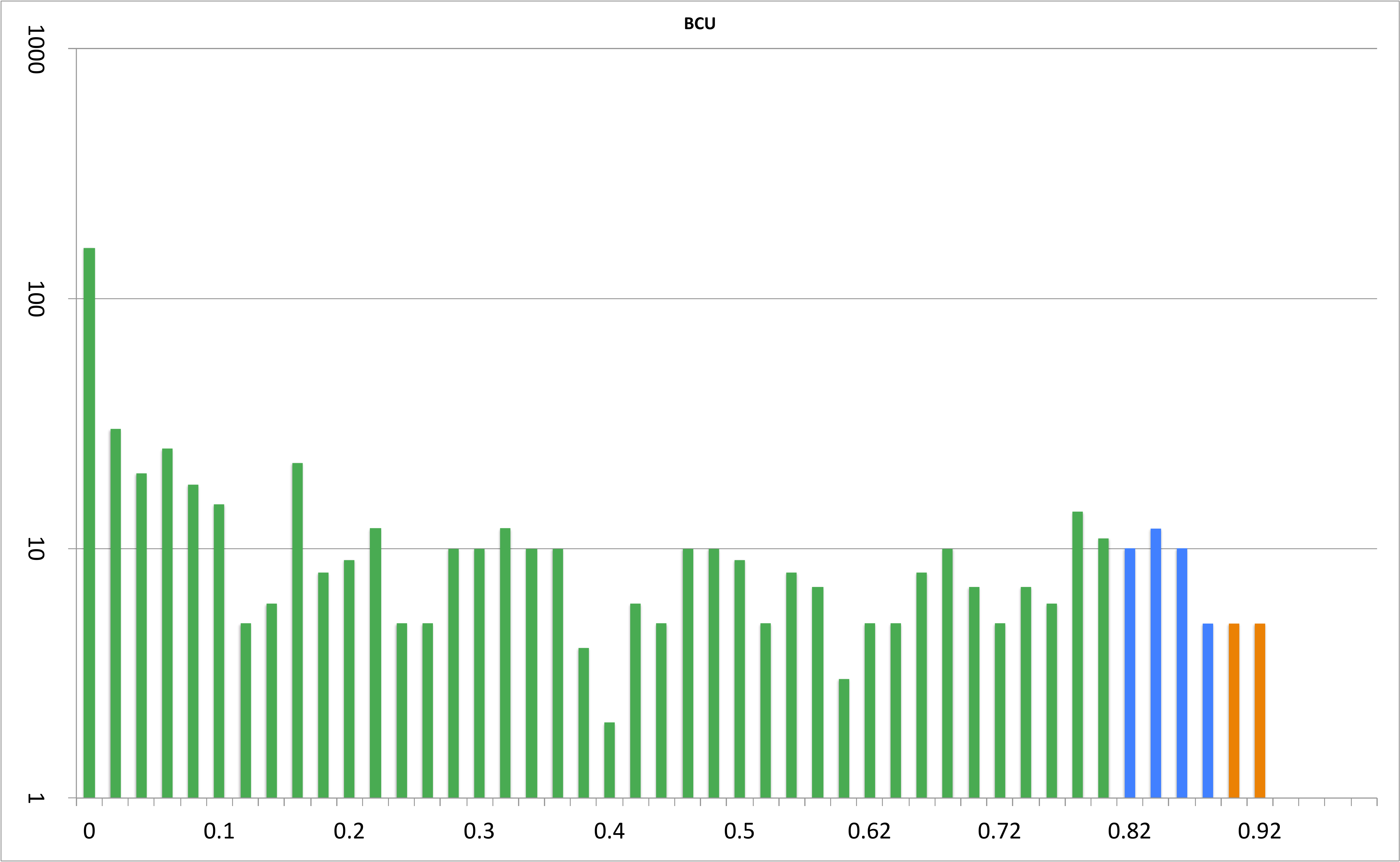} 
\includegraphics[width=0.48 \textwidth]{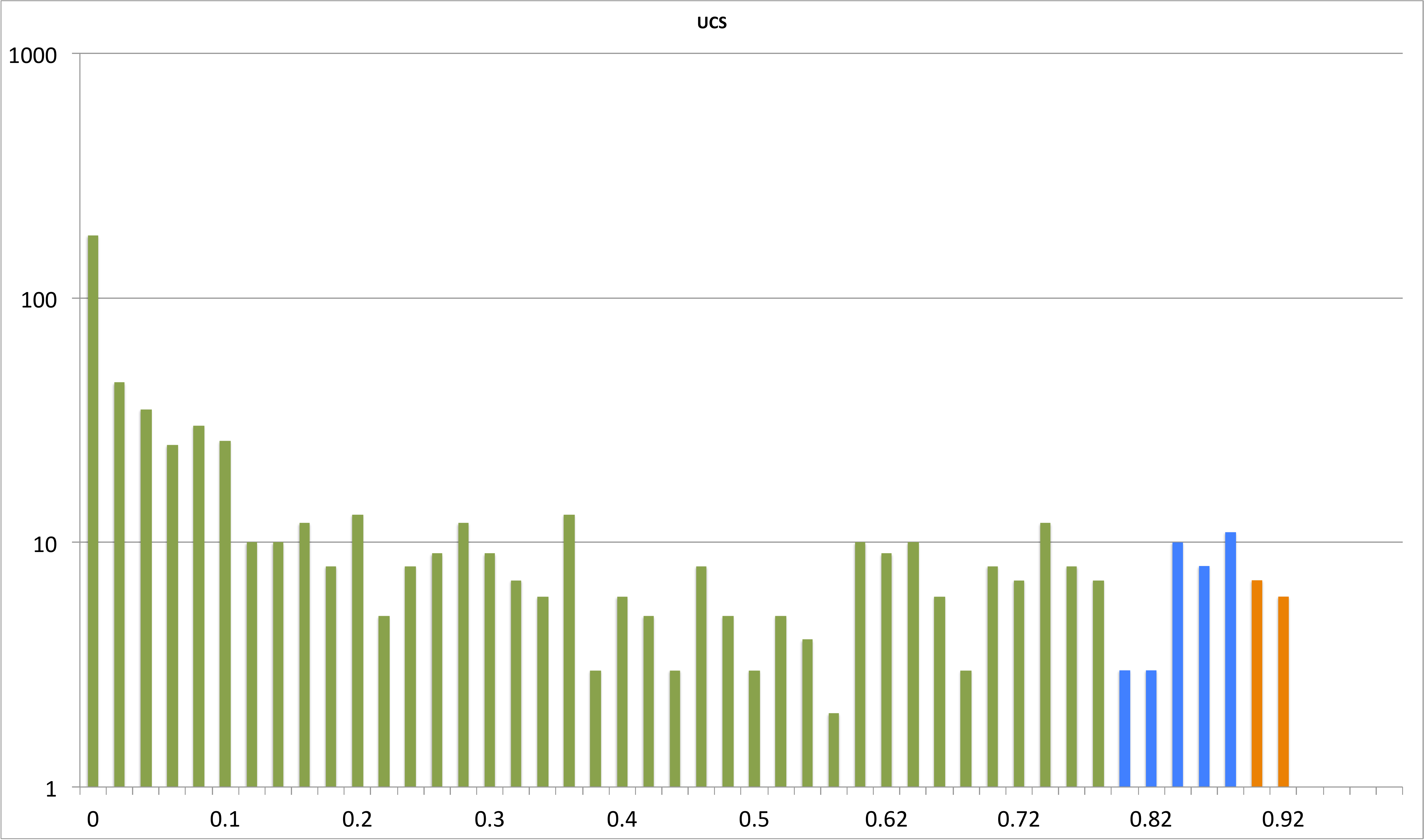}
\caption{\label{can} ANN likelihood to be HSP candidates of 3FGL BCUs (left) and UCS$_{agn}$  (right). 
%Vertical blue and steel blue lines indicate the classification thresholds of our ANN algorithm to identify sources with a  $L_{HSP} > 0.8$ and VHC candidates with $L_{HSP} > 0.89$. 
Blue bars: sources with  $0.89 >L_{HSP} > 0.8$; Red bars: VHC candidates with $L_{HSP} > 0.89$}
\end{figure*}

We then applied the %optimized 
 ANN HSP algorithm to the 573 BCUs and the 559 UCS$_{agn}$ of our sample. %The likelihood distribution of both groups of sources is shown in Fig.~\ref{can}. 
 The resulting likelihood distributions plotted in Fig.~\ref{can} show, as expected, a  peak at $L_{HSP}$ = 0.0 due to the non-HSP populations (ISP and LSP), which are much more numerous than HSPs at energies covered by the {\it Fermi} LAT.
As for the original analysis \citep{bflap}, the lack of a peak at $L_{HSP}$ = 1.0 
%due the ECDF of the blazar subclasses, 
indicates that the ANN network was not able to separate HSPs cleanly, but for the purpose of selecting candidates for additional analysis we are primarily interested in finding a high fraction of the sources with the desired characteristics.   %on the
%basis of ECDF , however 
Requiring  the $L_{HSP}$ $>$ 0.8 value, we identified 48 BCU and  32 UCS$_{agn}$ as HSP candidates. 
%In order to have the cleanest candidates we define an even narrower   
Applying the higher threshold value $L_{HSP}$ $>$ 0.89, 
%(true positive ratio $\sim{90}\%$) 
% limits the non-contamination area in Fig.~\ref{can}. Applying this further threshold, 
we identified  11 BCUs and 5 UCS$_{agn}$ as Very High Confidence (VHC) HSP candidates. 
Table 1 and Table 2 show the full lists of candidates.  The names of the VHC candidates are shown in bold. 

 We compared the results of our analysis with the list of blazars presented in the 2WHSP catalog \citep{2wHSP}, since both methods attempt to identify HSP objects.  Of the 11 VHC candidates from the BCU list, 6 were also identified by 2WHSP: 3FGL J0506.9$-$5435, 3FGL J0921.0$-$2258, 3FGL J1155.4$-$3417, 3FGL J1711.6+8846, 3FGL J1714.1$-$2029, and 3FGL J1944.1$-$4523.  None of the 5 VHC candidates from the UCS list were found in the 2WHSP catalog.  Similar fractions were found for the other parts of the lists, indicating that the two methods are complementary in finding candidate HSPs.   
%, where the VHC sources are on the top of the list. \\

%In Figure~\ref{3d} we show the properties of the HC candidates (black dots) compared to those of the 3FGL  blazar subclasses HSP (blue), LSP (red), and ISP (green) in a 3D plot: $\gamma$-ray flux, Variability Index, and spectral index. All the candidates lie in a clean HSP area of the plot and far from the LSPs (red). 
%We also compare the HSP candidates with the 2WHSP catalogue \citep{2wHSP}, and 36 sources in Table 1 and Table 2 are also in the 2WHSP catalogue. Those results reinforce the ability of the algorithm to select consistent candidates.

%\begin{figure*}
%\begin{center}
%\includegraphics[width=0.9 \textwidth]{3dok.pdf}
%\caption{\label{3d} HC candidate (black dots) properties compared to those of  the 3FGL blazar subclasses HSP (blue), IBL (green), and LBL (red). 
%} 
%\end{center}
%\end{figure*}

\section {Fermi-LAT Spectral Analysis}
\label{<4>}
 Because the 4FGL catalog will soon be available with a larger sample of sources to study, we chose to focus our spectral analysis for this new method of selecting HSP candidates on the 16 VHC sources.
We analyzed 104 months of {\it Fermi}-LAT Pass 8 \citep{P8} data 
%We performed an analysis of {\it {Fermi}}-LAT data analyzing 104 months of Pass 8 data, \citep{P8}  
from 2008 August 4 to 2017 April 4, selecting $\gamma$-ray events in the energy range $E=[0.1,1000]$ GeV, passing standard data quality selection criteria and zenith angle cuts for AGN \citep[e.g.][]{meyer},
in order to find the $\gamma$-ray properties of our HSP candidates. 
We considered events belonging to the Pass~8 {\tt SOURCE} event class and used the corresponding  instrument response functions {\tt P8R2\_SOURCE\_V6}, since we were interested in point source detection. 
We used the interstellar emission model (IEM) released with Pass 8 data \citep{Acero:2016qlg} (i.e., {\tt gll\_iem\_v06.fits}).
This is the model recommended for use with Pass 8 analyses. 
We also included the standard template for the isotropic emission ({\tt iso\_P8R2\_SOURCE\_V6\_v06.txt})\footnote{For descriptions of these templates, see \url{http://fermi.gsfc.nasa.gov/ssc/data/access/lat/BackgroundModels.html}.}.

We developed an analysis pipeline using {\tt FermiPy}, a Python package that automates analyses with the {\it Fermi} Science Tools \citep{2017arXiv170709551W}\footnote{See \url{http://fermipy.readthedocs.io/en/latest/}.}.
{\tt FermiPy} includes tools that can 1) generate simulations of the $\gamma$-ray sky, 2) detect point sources, and 3) calculate the characteristics of their SEDs.   
%For more details on {\tt FermiPy} we refer to the Appendices of \citet{Fermi-LAT:2017yoi}. 

%The {\it{Fermi}} Science Tools work initially with an input related to the error circle, then data is binned in longitude and latitude and energy.
The likelihood analysis works on a square region of interest (ROI). We used a $16^{\circ}\times16^{\circ}$ ROI centered on the sources of our sample. 
We analyzed each ROI separately.
In each ROI, we binned the data with a pixel size of $0.08^{\circ}$ and 8 energy bins per decade.
Our background model included the IEM, isotropic template and sources from the preliminary 8-year list, FL8Y\footnote{\url{https://fermi.gsfc.nasa.gov/ssc/data/access/lat/fl8y/gll_psc_8year_v5.fit}}, except for the source being analyzed.
We allowed the normalization and slope of the IEM and the normalization of the isotropic template  to vary.
We first relocalized the source of interest, and then we searched for new point sources with Test Statistic $TS >25$, defined as twice the difference between the log-likelihood for the null hypothesis (no source) and the hypothesis of a source at the location.  New sources were then added to the analysis. We then analyzed each of our HSP candidates, assuming a power-law spectrum after determining that all the VHC candidates used that spectral form in the 3FGL catalog \citep{3FGL}. For the individual energy bins, we plotted as upper limits those points with a $TS <9$.
%We apply a power-law SED for the sources in our sample.
%After this first step we calculate the SED of the source and we compute its lightcurve in order to determine whether the source is variable. The variability was estimated for each source by calculating the Test Statistic for variability ($TS_{VAR}$), defined as twice the difference between the log-likelihood for the null hypothesis (constant flux in time) and the hypothesis of a variable flux.

In Table 1 and Table 2 we report the following parameters for the HSP candidates: the best fit and $1\sigma$ error of the position, the $TS$ of detection,  the photon index found for a power-law SED shape, the flux, and the value of $L_{HSP}$. 
%and energy flux integrated between 100 MeV to 300 GeV. 
The photon index parameter, if less than 2, can be a relevant indicator for an IC peak at TeV energies and therefore quite useful in selecting IACT and CTA  candidates. The mean and rms of the photon indexes of HSPs are $1.87\pm0.20$ while for LSPs and ISPs these are $2.21\pm0.18$, $2.07\pm0.20$ respectively \citep{3lac}. 

We have repeated the analysis with a log-parabola intrinsic spectrum instead of a power law. Only for three sources (3FGL~J0921.0-2258, 3FGL~J0153.4+7114, and 3FGL~J0506.9-5435) did we find that a log-parabola is preferred at the 4\,$\sigma$ level, i.e. $\mathrm{TS}_\mathrm{curv} > 16$, where $\mathrm{TS}_\mathrm{curv}$ is twice the difference of the log-likelihood values of the best fit with a log-parabola and a power law, respectively.
These results are largely consistent with the spectral parameters listed in the preliminary 4FGL catalog\footnote{\url{https://fermi.gsfc.nasa.gov/ssc/data/access/lat/8yr_catalog/}}.  The 4FGL sources corresponding to 3FGL J0921.0$-$2258 and 3FGL J0153.4+7114 have power-law spectra instead of the curved spectra we found, although the evidence for curvature is of low significance in our analysis. If we had used the power-law fit, the extrapolations would have obviously been higher.
%The only one of our VHC candidates that shows a preferred log-parabola spectral shape is the counterpart of 3FGL~J0506.9$-$5435.

\section{TeV candidates}
\label{<5>}

In this section, we compare the extrapolated fluxes of the {\it Fermi}-LAT spectra of our VHC sources to the sensitivity of present IACTs and the future CTA. 
%We use the {\it Fermi}-LAT spectra obtained in the previous section and focus further analysis on the VHC candidates. 
In the case that the log-parabola fit to the {\it Fermi}-LAT data is preferred with ${\rm TS}_{\rm curv} > 16$, we use the log-parabola parametrization for the extrapolation and the best-fit power law otherwise.
In order to evaluate whether the VHC HSP candidates can realistically be observed with IACTs or CTA, we must take into account the interaction of $\gamma$-rays with photons of  the extragalactic background light (EBL). The EBL spans the wavelength regime from ultraviolet to far-infrared wavelengths and mainly consists of the integrated starlight emitted over the history of the Universe and starlight  absorbed and re-emitted by dust in galaxies \citep{hauser,kashlinsky}.
During the propagation of  $\gamma$ rays through the EBL, the electron-positron pairs produced via $\gamma\gamma\to e^+e^-$ 
leads to an attenuation of the initial $\gamma$-ray flux \citep{nikisov,gould,dwe}. 
To properly evaluate the absorption effect of the EBL it is necessary to know the redshift of the analyzed $\gamma$-ray source. Since the redshifts of the selected VHC HSP candidates are unknown, we assume redshifts  of $z = 0$ and $z = 0.5$ for the calculation. These are typical values of observed $\gamma$-ray BL Lacs\footnote{We recognize that a redshift identically 0 is unrealistic. This choice provides a ``best case'' limit.}. 
We use these $z$ values while acknowledging that the blazar redshift range is a very open and long-standing debate. Some authors argue that the BL Lacs without a redshift are likely much more distant than those with a measured one \citep[e.g.,][]{pad}, so it could be possible that some objects fall beyond the 0.5 redshift.

\begin{figure}[p]
\centering
\includegraphics[width=0.9\textwidth]{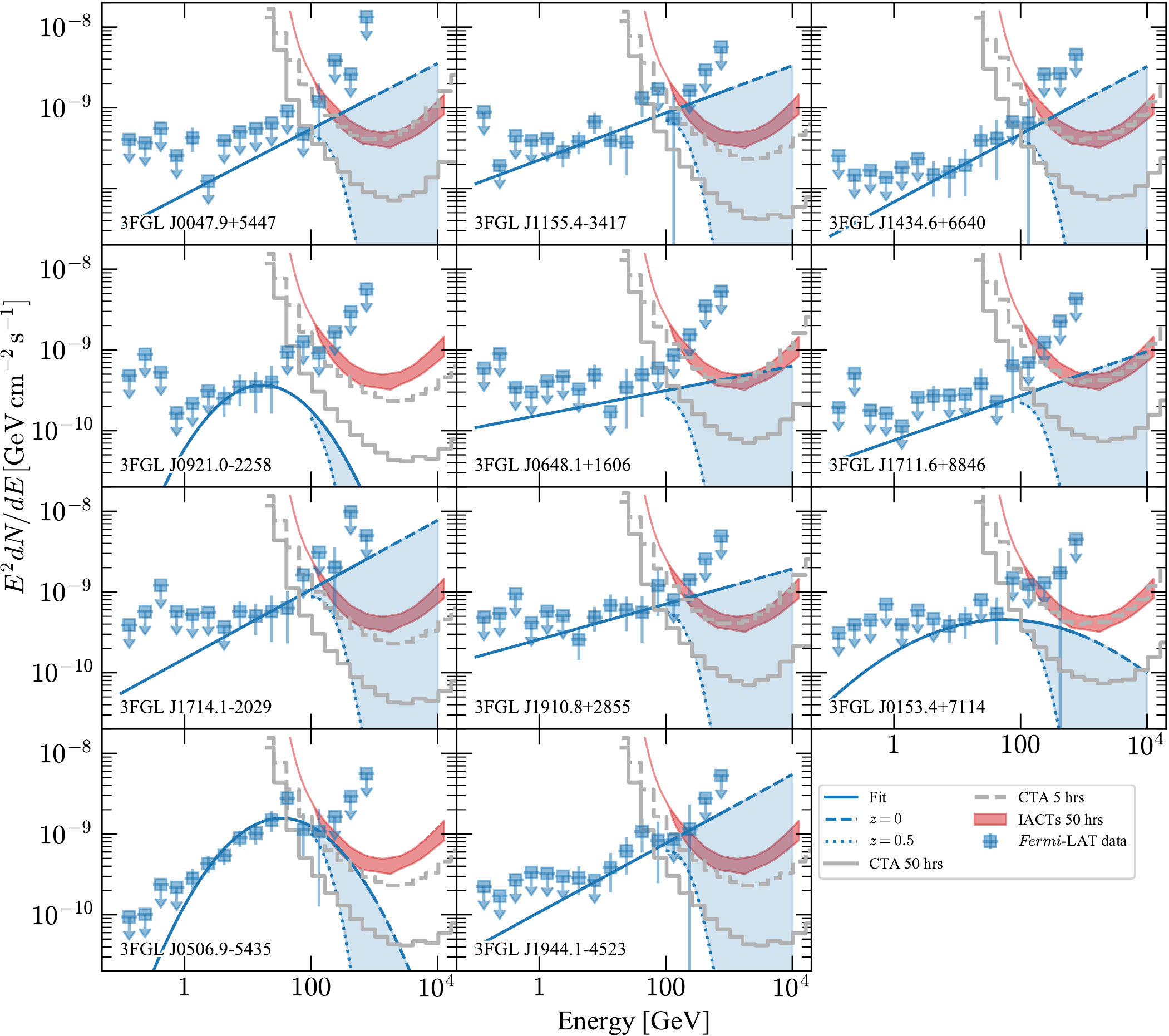}
\caption{\label{bcu}SEDs of the VHC BCU sources as TeV candidates. The dashed (dotted) line denotes the extrapolation of the best-fit {\it Fermi}-LAT spectra up to 10\,TeV for a redshift of $z = 0$ ($z = 0.5$) The shaded region indicates the possible source flux for redshifts between $0 < z \leq 0.5$. The CTA  sensitivity for 5 (50) hours of observation is shown as a gray dashed (solid) line. The sensitivity curve for either the northern or southern array is used as described in the main text. The 50 hour sensitivity for currently operating IACTs is shown as a red shaded band.}
\end{figure}

\begin{figure}[p]
\centering
\includegraphics[width=0.9\textwidth]{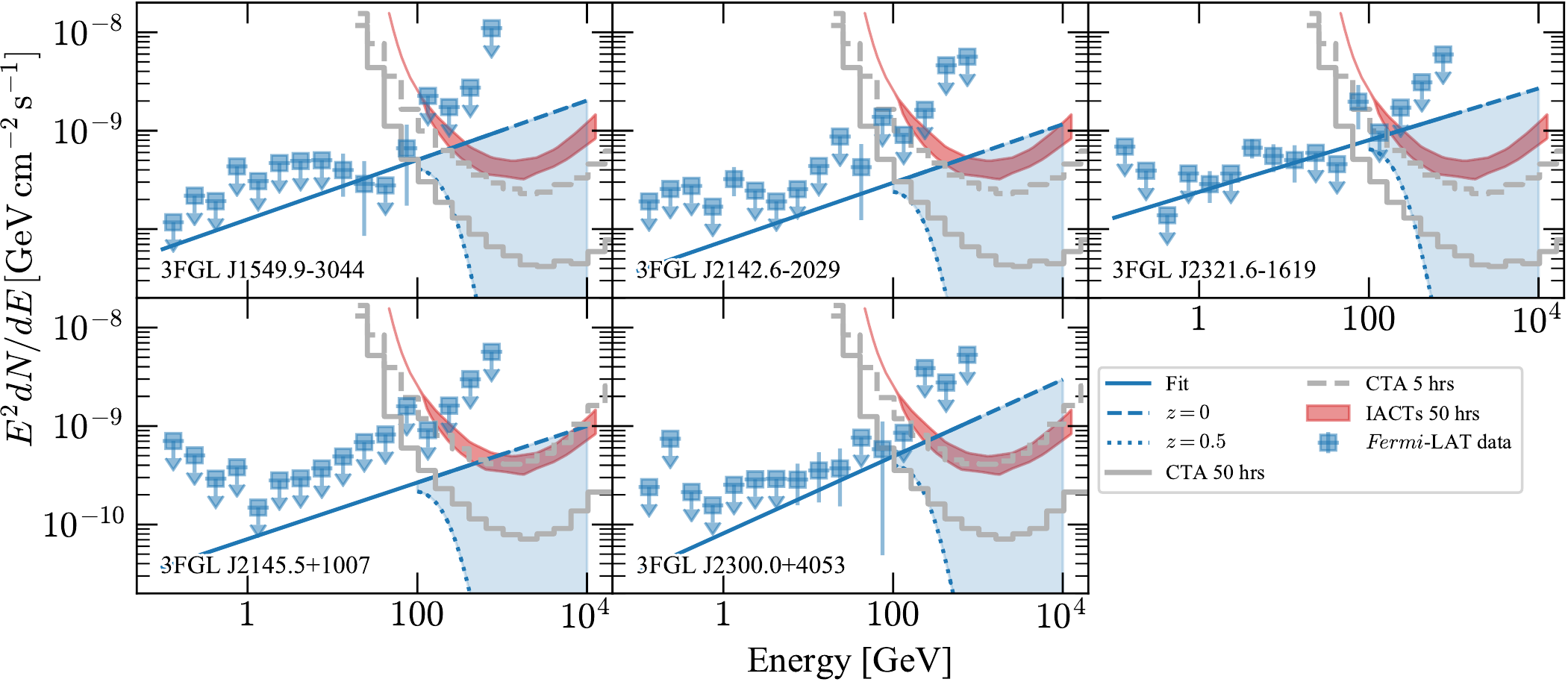}
\caption{\label{ucscta}SEDs of the VHC UCS sources as TeV candidates. The dashed (dotted) line denotes the extrapolation of the best-fit {\it Fermi}-LAT spectra up to 10\,TeV for a redshift of $z = 0$ ($z = 0.5$) The shaded region indicates the possible source flux for redshifts between $0 < z \leq 0.5$. The CTA  sensitivity for 5 (50) hours of observation is shown as a gray dashed (solid) line. The sensitivity curves for either the northern of southern array is used as described in the main text. The 50 hour sensitivity for currently operating IACTs is shown as a red shaded band. }
\end{figure}

Using the EBL model of \citet{dominguez}, we extrapolate the best-fit spectra obtained with the \emph{Fermi}-LAT analysis in Sec.~\ref{<4>} up to 10\,TeV with the assumed redshift values. The results are shown in Fig.~\ref{bcu} and Fig.~\ref{ucscta} where we compare the extrapolated fluxes with the CTA sensitivity~\citep{ctasens} for 50\,hours (5\,hours) of observations as a solid (dashed) gray line.
The sensitivities for currently operating IACT arrays H.E.S.S., MAGIC, and VERITAS are shown as a red band \citep{hessperf,magicperf}\footnote{The VERITAS sensitivity curve from \url{https://veritas.sao.arizona.edu/about-veritas-mainmenu-81/veritas-specifications-mainmenu-111} is used.}. 
The CTA sensitivity curves are available for the northern and southern array and for zenith angles ($Z$) of $20^\circ$ and $40^\circ$. We choose the sensitivity curve depending on the source declination $\delta$ assuming CTA site locations at 28.7${}^\circ$ northern latitude and 24.7${}^\circ$ southern latitude. For $\delta > 58.7^\circ$,  we choose the northern array with $Z = 40^\circ$, for $58.7^\circ \leqslant \delta < 2.7^\circ$ the northern array with $Z=20^\circ$, for $2.7 \geqslant Z \geqslant -54.7^\circ$ the southern array with $Z = 20^\circ$, and the southern array with $Z=40^\circ$ for $\delta < -54.7^\circ$\footnote{The sensitivity curves of the northern and southern array  are available at \url{www.cta-observatory.org}}.

 In Table~\ref{tab:zmax},  we report an estimate of the maximum redshift values of the VHC TeV candidates so that the sources are still detectable at 5\,$\sigma$ for 50-hr observations with current IACTs as well as 5- and 50-hr CTA  observations.
 For these estimates, we assume that the spectra are detectable if the extrapolation is above the sensitivity curves. 
 We also report how these values change if the best-fit spectral parameters are changed by their $1~\sigma$ statistical uncertainties. 
 An actual detection will depend on the spectral shape and the exact exposure time, and therefore these numbers should be regarded as a rough estimate. 
 If no value is given, the source might not be significantly detected with the assumed observation within the assumed redshift range. 
 
 Any of the VHC sources that lie at low redshift should be detectable with CTA, with the exception of  3FGL~J0921.0$-$2258, which shows strong intrinsic curvature.
 Similarly, most sources should already be detectable by the present-generation IACTs if their redshifts are close to 0. 
 For a redshift of $z = 0.5$, most sources appear to be good candidates for detection with 50 hours of CTA observations, primarily in the 100-GeV energy range.  Only 3FGL~J1714.1$-$2029 seems possible as a candidate for current IACTs if its distance is at the upper end of the range considered, due to the EBL attenuation. 

\section{Conclusions}
\label{<6>}
Motivated to expand the sample of TeV sources, %we have developed a search method based on machine learning 
we applied a machine learning algorithm  to variability parameters of {\it {Fermi}}-LAT blazar-like sources without firm identifications combined with new analysis of the LAT data for these sources.  Follow-up work will require additional multiwavelength studies, including finding redshifts for most of the candidates and targeted observation by IACTs.  We also recognize that this search is necessarily incomplete because of the difficulty to distinguish the blazar subclasses  by the $\gamma$-ray flux properties only. As already pointed out by \citet{3lac}, the $\gamma$-ray sources with unknown properties are generally fainter than the well-defined classes. The fainter sources offer less of the flaring information needed for the machine learning method, and so there may be HSP blazars among the sample of 3FGL sources rejected in the first step of our method. The level of incompleteness is difficult to quantify due to the very similar values of the synchrotron peaks of the three blazar subclasses. Nevertheless, some VHC HSP candidates, also thanks to the analysis of {\it {Fermi}}-LAT data, are convincing as TeV candidates and should be promising targets for currently operating IACTs, especially if the sources are located below reshifts of $\sim 0.1$. If the sources are further away, CTA should be capable of significantly detecting them.

\acknowledgments

The \textit{Fermi} LAT Collaboration acknowledges generous ongoing support
from a number of agencies and institutes that have supported both the
development and the operation of the LAT as well as scientific data analysis.
These include the National Aeronautics and Space Administration and the
Department of Energy in the United States, the Commissariat \`a l'Energie Atomique
and the Centre National de la Recherche Scientifique / Institut National de Physique
Nucl\'eaire et de Physique des Particules in France, the Agenzia Spaziale Italiana
and the Istituto Nazionale di Fisica Nucleare in Italy, the Ministry of Education,
Culture, Sports, Science and Technology (MEXT), High Energy Accelerator Research
Organization (KEK) and Japan Aerospace Exploration Agency (JAXA) in Japan, and
the K.~A.~Wallenberg Foundation, the Swedish Research Council and the
Swedish National Space Board in Sweden. Additional support for science analysis during the operations phase 
is gratefully acknowledged from the Istituto Nazionale di Astrofisica in Italy and the Centre
National d'\'Etudes Spatiales in France. This work performed in part under DOE
Contract DE-AC02-76SF00515.

MDM acknowledges support by the NASA {\it Fermi} Guest Investigator Program 2014 through the {\it Fermi} multi-year Large Program No. 81303 (P.I. E.~Charles) and by the NASA {\it Fermi} Guest Investigator Program 2016 through the {\it Fermi} one-year Program No. 91245 (P.I. M.~Di Mauro). 

This  paper  has  gone  through  internal  review  by  the CTA  Consortium.
This research has made use of the CTA instrument response functions provided by the CTA Consortium and Observatory, see \url{http://www.cta-observatory.org/science/cta-performance/} (version prod3b-v1) for more details.
\vspace{5mm}
\facilities{{\it Fermi}, H.E.S.S., MAGIC, VERITAS, CTA}
\begin{table}
\clearpage
\begin{small}
\caption{\label{table1}Results of the {\it Fermi}-LAT analysis of the full list of BCU HSP candidates. Names of the high-confidence sources are shown in {\bf bold}. Columns: (1) Source name; (2) RA; (3) DEC; (4) $68\%$ error on the position $\theta$;  (5) Detection TS 0.1-- 300 GeV;  (6)   Photon Index; (7)   Integrated flux in the energy range 0.1-- 300 GeV;  8) $L_{HSP}$, likelihood the source is an HSP   }
\begin{tabular}{lcccccccc}
\hline
\hline
Source name  & RA   &  DEC  &  $\theta_{68\%}$	& TS  & Ph.Index   &  Flux &  $L_{HSP}$\\
\hline
            & [deg] & [deg]  & [deg]  &  &   & [$10^{-9}$ ph/cm$^2$/s] & \\
\hline
\hline
3FGL J0030.2$-$1646  &   7.59  &   -16.82   &   0.02   &   168.7 &   $1.66\pm0.08$   &   $13.4\pm3.3$ & 0.80  \\
3FGL J0039.0$-$2218  &   9.77  &   -22.32   &   0.03   &   89.3 &   $1.67\pm0.11$   &   $9.2\pm2.9$ &  0.86 \\
3FGL J0040.3+4049  &   10.09  &   40.83   &   0.03   &   75.9 &   $1.92\pm0.16$   &   $18.2\pm9.4$ & 0.87  \\
3FGL J0043.5$-$0444  &   10.88  &   -4.72   &   0.04   &   54.0 &   $1.91\pm0.15$   &   $16.2\pm7.2$ & 0.83  \\
3FGL J0043.7$-$1117  &   10.94  &   -11.31   &   0.04   &   69.4 &   $1.86\pm0.12$   &   $16.0\pm5.3$ & 0.88  \\
{\bf 3FGL J0047.9+5447}  &   12.02  &   54.81   &   0.03   &   56.7 &   $1.58\pm0.17$   &   $4.9\pm3.2$ & 0.92  \\
3FGL J0132.5$-$0802  &   23.19  &   -8.07   &   0.03   &   71.9 &   $1.87\pm0.11$   &   $16.8\pm5.4$ & 0.84  \\
{\bf 3FGL J0153.4+7114}  &   28.43  &   71.26   &   0.02   &   80.9 &   $1.82\pm0.13$   &   $17.5\pm7.3$ &  0.89 \\
3FGL J0204.2+2420  &   31.09  &   24.27   &   0.04   &   27.6 &   $1.70\pm0.16$   &   $4.7\pm2.6$ &  0.81 \\
3FGL J0305.2$-$1607  &   46.29  &   -16.14   &   0.02   &   147.6 &   $1.80\pm0.11$   &   $17.8\pm5.8$ &  0.86 \\
3FGL J0342.6$-$3006  &   55.71  &   -30.11   &   0.04   &   43.2 &   $1.96\pm0.14$   &   $12.5\pm4.7$ & 0.84  \\
3FGL J0439.6$-$3159  &   69.85  &   -32.03   &   0.03   &   119.9 &   $1.75\pm0.09$   &   $13.3\pm3.9$ & 0.81  \\
{\bf 3FGL J0506.9$-$5435}  &   76.76  &   -54.60   &   0.01   &   455.4 &   $1.50\pm0.05$   &   $14.2\pm2.2$ & 0.89  \\
3FGL J0515.5$-$0123  &   78.87  &   -1.42   &   0.04   &   45.7 &   $1.80\pm0.11$   &   $11.4\pm4.4$ & 0.85  \\
3FGL J0528.3+1815  &   82.11  &   18.27   &   0.04   &   35.7 &   $1.67\pm0.15$   &   $6.6\pm3.6$ & 0.87  \\
3FGL J0620.4+2644  &   95.17  &   26.74   &   0.02   &   92.0 &   $1.54\pm0.11$   &   $6.3\pm2.8$ & 0.85  \\
3FGL J0640.0$-$1252  &   100.01  &   -12.90   &   0.02   &   174.1 &   $1.52\pm0.09$   &   $10.3\pm3.4$ &  0.85 \\
3FGL J0646.4$-$5452  &   101.62  &   -54.92   &   0.03   &   190.3 &   $1.46\pm0.29$   &   $8.8\pm1.8$ & 0.87  \\
{\bf 3FGL J0648.1+1606}  &   102.03  &   16.09   &   0.03   &   40.1 &   $1.82\pm0.16$   &   $10.7\pm5.9$ & 0.90  \\
3FGL J0650.5+2055  &   102.64  &   20.93   &   0.02   &   206.2 &   $1.72\pm0.08$   &   $21.9\pm5.7$ & 0.82  \\
3FGL J0733.5+5153  &   113.35  &   51.86   &   0.03   &   104.3 &   $1.69\pm0.10$   &   $9.9\pm3.2$ &  0.85 \\
3FGL J0742.4$-$8133c  &   115.45  &   -81.54   &   0.05   &   32.3 &   $2.03\pm0.28$   &   $21.1\pm15.4$ &  0.88 \\
3FGL J0746.9+8511  &   117.25  &   85.22   &   0.03   &   119.0 &   $1.68\pm0.09$   &   $10.0\pm2.8$ & 0.83  \\
{\bf 3FGL J0921.0$-$2258}  &   140.24  &   -22.95   &   0.03   &   62.5 &   $1.74\pm0.14$   &   $9.4\pm4.1$ & 0.91  \\
3FGL J1040.8+1342  &   160.26  &   13.72   &   0.03   &   69.1 &   $1.71\pm0.13$   &   $8.3\pm3.4$ & 0.86  \\
3FGL J1141.2+6805  &   175.33  &   68.08   &   0.02   &   140.1 &   $1.69\pm0.09$   &   $10.9\pm2.8$ & 0.85  \\
{\bf 3FGL J1155.4$-$3417}  &   178.87  &   -34.33   &   0.02   &   147.3 &   $1.64\pm0.09$   &   $11.8\pm3.3$ & 0.92  \\
3FGL J1158.9+0818  &   179.71  &   8.31   &   0.04   &   51.5 &   $1.81\pm0.14$  &   $11.0\pm4.6$ & 0.80  \\
3FGL J1203.5$-$3925  &   180.85  &   -39.42   &   0.03   &   103.2 &   $1.70\pm0.10$   &   $13.5\pm4.5$ & 0.85  \\
3FGL J1319.6+7759  &   199.95  &   78.01   &   0.02   &   182.6 &   $1.95\pm0.8$   &   $28.3\pm5.9$ & 0.82  \\
{\bf 3FGL J1434.6+6640}  &   218.72  &   66.67   &   0.03   &   73.9 &   $1.58\pm0.12$   &   $4.4\pm1.7$ & 0.92  \\
3FGL J1446.8$-$1831  &   221.75  &   -18.51   &   0.05   &   27.9 &   $1.71\pm0.15$   &   $6.1\pm3.3$ &  0.84 \\
3FGL J1547.1$-$2801  &   236.81  &   -28.04   &   0.03   &   96.8 &   $1.78\pm0.10$   &   $19.0\pm6.1$ & 0.81  \\
3FGL J1612.4$-$3100  &   243.10  &   -30.99   &   0.02   &   495.0 &   $1.86\pm0.08$   &   $38.0\pm7.8$ & 0.81  \\
{\bf 3FGL J1714.1$-$2029}  &   258.52  &   -20.48   &   0.03   &   73.8 &   $1.44\pm0.12$   &   $5.1\pm2.3$ & 0.90  \\
{\bf 3FGL J1711.6+8846}  &   258.67  &   88.75   &   0.04   &   44.3 &   $1.83\pm0.15$   &   $8.8\pm4.2$ & 0.90  \\
3FGL J1824.4+4310  &   276.12  &   43.18   &   0.03   &   80.9 &   $1.83\pm0.15$   &   $13.6\pm5.3$ & 0.88  \\
3FGL J1841.2+2910  &   280.36  &   29.16   &   0.02   &   195.9 &   $1.80\pm0.08$   &   $29.0\pm7.1$ & 0.80  \\
3FGL J1855.1$-$6008  &   283.67  &   -60.13   &   0.06   &   21.4 &   $1.84\pm0.16$   &   $7.2\pm3.9$ &  0.84 \\
3FGL J1908.8$-$0130  &   287.20  &   -1.53   &   0.02   &   306.4 &   $1.52\pm0.21$   &   $18.6\pm2.8$ & 0.82  \\
{\bf 3FGL J1910.8+2855}  &   287.71  &   28.94   &   0.02   &   102.3 &   $1.62\pm0.10$   &   $9.8\pm3.3$ &  0.90 \\
3FGL J1939.6$-$4925  &   294.96  &   -49.47   &   0.03   &   64.6 &   $1.85\pm0.11$   &   $14.9\pm5.0$ &  0.85 \\
{\bf 3FGL J1944.1$-$4523}  &   296.11  &   -45.38   &   0.02   &   100.7 &   $1.64\pm0.10$   &   $9.4\pm3.3$ & 0.89  \\
3FGL J1959.8$-$4725  &   299.94  &   -47.43   &   0.01   &   923.8 &   $1.52\pm0.08$   &   $30.5\pm5.0$ & 0.87  \\
3FGL J2108.6$-$8619  &   316.99  &   -86.31   &   0.03   &   91.0 &   $1.65\pm0.12$   &   $10.3\pm4.0$ & 0.87  \\
3FGL J2312.9$-$6923  &   348.40  &   -69.39   &   0.04   &   35.3 &   $1.72\pm0.17$   &   $5.5\pm2.9$ &  0.86 \\
3FGL J2316.8$-$5209  &   349.28  &   -52.19   &   0.06   &   37.3 &   $1.89\pm0.16$   &   $10.8\pm5.0$ & 0.85  \\
3FGL J2347.9+5436  &   356.97  &   54.58   &   0.02   &   163.0 &   $1.79\pm0.08$   &   $24.2\pm6.4$ & 0.82  \\
\hline
\end{tabular}
\end{small}
\label{<Table 1>}
\end{table} 
%\end{landscape}

%\begin{landscape}
\begin{table}
\clearpage
\begin{small}
\caption{Same as Table~\ref{<Table 1>} for UCS$_{agn}$ HSP candidates.}
\begin{tabular}{lcccccccc}
\hline
\hline
Source name  & RA   &  DEC  &  $pos_{68}\%$	& TS  & Ph.Index   &  Flux &  $L_{HSP}$\\
\hline
            & [deg] & [deg]  & [deg]  &  &   & [$10^{-9}$ ph/cm$^2$/s] & \\
\hline
\hline
3FGL J0020.9+0323  &   5.26  &   3.36   &   0.04   &   60.7 &   $2.01\pm0.14$   &   $23.3\pm8.4$ &  0.88 \\
3FGL J0049.0+4224  &   12.26  &   42.38   &   0.04   &   37.0 &   $1.81\pm0.16$   &   $8.1\pm4.4$ &  0.82 \\
3FGL J0234.2$-$0629  &   38.59  &   -6.47   &   0.03   &   90.7 &   $1.83\pm0.11$   &   $15.6\pm4.8$ &  0.84 \\
3FGL J0251.1$-$1829  &   42.79  &   -18.50   &   0.02   &   104.3 &   $1.59\pm0.10$   &   $7.0\pm2.2$ & 0.88  \\
3FGL J0312.7$-$2222  &   48.15  &   -22.36   &   0.02   &   177.1 &   $1.84\pm0.08$   &   $22.3\pm5.2$ & 0.87  \\
3FGL J0506.9+0321  &   76.71  &   3.38   &   0.03   &   77.1 &   $1.81\pm0.12$   &   $15.3\pm6.0$ & 0.89  \\
3FGL J0524.5$-$6937  &   81.16  &   -69.61   &   0.03   &   94.1 &   $2.05\pm0.15$   &   $49.4\pm21.2$ & 0.86  \\
3FGL J0527.3+6647  &   81.86  &   66.80   &   0.03   &   51.9 &   $1.91\pm0.15$   &   $13.0\pm6.0$ & 0.83  \\
3FGL J0731.8$-$3010  &   112.96  &   -30.13   &   0.04   &   37.1 &   $1.96\pm0.17$   &   $22.4\pm11.9$ &  0.84 \\
3FGL J0813.5$-$0356  &   123.45  &   -3.95   &   0.04   &   57.0 &   $1.71\pm0.12$   &   $9.1\pm3.5$ & 0.88  \\
3FGL J0928.3$-$5255  &   142.09  &   -52.94   &   0.02   &   98.7 &   $2.09\pm0.09$   &   $88.0\pm25.5$ & 0.80  \\
3FGL J0952.8+0711  &   148.22  &   7.23   &   0.04   &   51.0 &   $1.92\pm0.15$   &   $14.0\pm6.0$ & 0.84  \\
3FGL J1057.6$-$4051  &   164.43  &   -40.87   &   0.03   &   40.2 &   $1.72\pm0.15$   &   $6.6\pm3.4$ & 0.82  \\
3FGL J1155.3$-$1112  &   178.82  &   -11.19   &   0.03   &   52.5 &   $2.03\pm0.15$   &   $21.2\pm8.9$ & 0.89  \\
3FGL J1222.7+7952  &   185.92  &   79.90   &   0.04   &   43.8 &   $2.13\pm0.22$   &   $21.1\pm12.1$ &  0.86 \\
3FGL J1225.4$-$3448  &   186.35  &   -34.75   &   0.05   &   22.3 &   $1.74\pm0.19$   &   $5.1\pm3.4$ &  0.86 \\
3FGL J1234.7$-$0437  &   188.71  &   -4.56   &   0.04   &   51.5 &   $2.01\pm0.14$   &   $23.5\pm9.7$ &  0.87 \\
3FGL J1513.3$-$3719  &   228.35  &   -37.39   &   0.03   &   54.7 &   $1.94\pm0.13$   &   $19.8\pm8.4$ &  0.87 \\
3FGL J1525.8$-$0834  &   231.53  &   -8.53   &   0.03   &   59.5 &   $1.92\pm0.12$   &   $20.0\pm7.3$ & 0.89  \\
3FGL J1528.1$-$2904  &   232.12  &   -29.11   &   0.06   &   26.3 &   $1.80\pm0.18$   &   $8.7\pm5.4$ & 0.83  \\
3FGL J1545.0$-$6641  &   236.21  &   -66.71   &   0.02   &   150.1 &   $1.60\pm0.10$   &   $11.2\pm3.9$ & 0.84  \\
{\bf 3FGL J1549.9$-$3044}  &   237.46  &   -30.75   &   0.02   &   64.3 &   $1.61\pm0.12$   &   $6.2\pm2.6$ & 0.91  \\
3FGL J1619.1+7538  &   244.78  &   75.61   &   0.02   &   107.1 &   $1.87\pm0.10$   &   $15.5\pm4.7$ & 0.88  \\
3FGL J1922.2+2313  &   290.57  &   23.25   &   0.03   &   80.8 &   $2.22\pm0.14$   &   $93.1\pm36.7$ & 0.87  \\
3FGL J2015.3$-$1431  &   303.81  &   -14.55   &   0.06   &   17.4 &   $1.81\pm0.21$   &   $5.6\pm4.2$ & 0.85  \\
3FGL J2043.6+0001  &   310.94  &   0.00   &   0.04   &   48.5 &   $2.02\pm0.15$   &   $21.5\pm8.1$ & 0.87  \\
3FGL J2053.9+2922  &   313.45  &   29.37   &   0.02   &   359.6 &   $1.77\pm0.06$   &   $46.0\pm8.4$ &  0.85 \\
{\bf 3FGL J2142.6$-$2029}  &   325.66  &   -20.50   &   0.04   &   36.1 &   $1.69\pm0.17$   &   $5.0\pm2.9$ & 0.91  \\
{\bf 3FGL J2145.5+1007}  &   326.38  &   10.13   &   0.03   &   34.1 &   $1.70\pm0.20$   &   $4.8\pm3.4$ & 0.90  \\
3FGL J2224.4+0351  &   336.12  &   3.89   &   0.05   &   29.5 &   $1.94\pm0.18$   &   $13.2\pm6.4$ & 0.89  \\
{\bf 3FGL J2300.0+4053}  &   345.06  &   40.88   &   0.03   &   52.5 &   $1.64\pm0.14$   &   $6.2\pm3.0$ &  0.90 \\
3FGL J2309.0+5428  &   347.20  &   54.41   &   0.03   &   77.1 &   $1.75\pm0.10$   &   $16.2\pm5.7$ &  0.85 \\
{\bf 3FGL J2321.6$-$1619}  &   350.40  &   -16.32   &   0.02   &   174.5 &   $1.73\pm0.08$   &   $17.3\pm4.1$ &  0.91 \\
\hline
\end{tabular}
\end{small}
\label{<Table 2>}
\end{table} 
%\end{landscape}
\begin{table*}
\clearpage
\begin{center}
\caption{\label{tab:zmax} Estimates of the maximum redshift values of the VHC sources for which 
the sources are still detectable at $5\,\sigma$ in a 50\,hours of current IACT and 50 (5)\,hours of CTA observations,
respectively.
If no value is given, the source will not be significantly detected within the assumed observation time.
Sub- and superscript numbers give the change in the redshift values if the \emph{Fermi}-LAT spectrum
is extrapolated within 1$\,\sigma$ uncertainties.
See main text for further details and caveats.}
\begin{tabular}{l|ccc}
\hline
\hline
Source name & $\mathrm{T}_\mathrm{obs} = 50\,\mathrm{hours}$, current IACTs & $\mathrm{T}_\mathrm{obs} = 5\,\mathrm{hours}$, CTA & $\mathrm{T}_\mathrm{obs} = 50\,\mathrm{hours}$, CTA\\
\hline
\multicolumn{4}{c}{BCU HC$_{TeV} $ candidates}\\
\hline
3FGL J0047.9+5447 &  $0.18^{>0.50}_{\text{---}}$ & $0.15^{>0.50}_{\text{---}}$ & $> 0.50_{-0.38}$ \\
3FGL J1155.4$-$3417 &  $0.26^{>0.50}_{-0.17}$ & $0.35^{>0.50}_{-0.24}$ & $> 0.50$ \\
3FGL J1434.6+6640 &  $0.15^{+0.27}_{-0.13}$ & $0.14^{+0.33}_{\text{---}}$ & $> 0.50_{-0.31}$ \\
3FGL J0921.0$-$2258 &  $\text{---}$ & $\text{---}$ & $\text{---}$ \\
3FGL J0648.1+1606 &  $0.03^{+0.19}_{\text{---}}$ & $0.01^{+0.19}_{\text{---}}$ & $0.29^{>0.50}_{-0.26}$ \\
3FGL J1711.6+8846 &  $0.04^{+0.22}_{\text{---}}$ & $0.02^{+0.24}_{\text{---}}$ & $0.32^{>0.50}_{-0.29}$ \\
3FGL J1714.1$-$2029 &  $0.38^{>0.50}_{-0.28}$ & $0.47^{>0.50}_{-0.35}$ & $> 0.50$ \\
3FGL J1910.8+2855 &  $0.18^{+0.25}_{-0.15}$ & $0.16^{+0.28}_{-0.15}$ & $> 0.50_{-0.17}$ \\
3FGL J0153.4+7114 &  $\text{---}$ & $\text{---}$ & $0.43^{>0.50}_{\text{---}}$ \\
3FGL J0506.9$-$5435 &  $0.05^{>0.50}_{\text{---}}$ & $0.25^{>0.50}_{\text{---}}$ & $> 0.50_{\text{---}}$ \\
3FGL J1944.1$-$4523 &  $0.27^{>0.50}_{-0.21}$ & $0.34^{>0.50}_{-0.26}$ & $> 0.50_{-0.07}$ \\

\hline
\multicolumn{4}{c}{UCS$_{agn}$ HC$_{TeV} $ candidates}\\
\hline
3FGL J1549.9$-$3044 &  $0.13^{+0.27}_{\text{---}}$ & $0.15^{>0.50}_{-0.13}$ & $> 0.50_{-0.29}$ \\
3FGL J2142.6$-$2029 &  $0.05^{+0.28}_{\text{---}}$ & $0.07^{+0.35}_{\text{---}}$ & $0.41^{>0.50}_{-0.35}$ \\
3FGL J2321.6$-$1619 &  $0.23^{+0.20}_{-0.15}$ & $0.30^{>0.50}_{-0.20}$ & $> 0.50$ \\
3FGL J2145.5+1007 &  $0.04^{+0.33}_{\text{---}}$ & $0.02^{+0.35}_{\text{---}}$ & $0.26^{>0.50}_{\text{---}}$ \\
3FGL J2300.0+4053 &  $0.15^{+0.32}_{\text{---}}$ & $0.13^{+0.35}_{\text{---}}$ & $> 0.50_{-0.37}$ \\
\hline
\end{tabular}
\end{center}
\end{table*}

%% This command is needed to show the entire author+affilation list when
%% the collaboration and author truncation commands are used.  It has to
%% go at the end of the manuscript.
%\allauthors

%% Include this line if you are using the \added, \replaced, \deleted
%% commands to see a summary list of all changes at the end of the article.
%\listofchanges

\end{document}